\documentstyle[amsmath,a4,11pt]{article}

\begin{document}
\title{An exact isotropic solution}

\author{A J John\footnote{Present address: Department of Mathematics and Applied Mathematics,
University of Cape Town, Rondebosch 7701, South Africa}
~and S D Maharaj\\
Astrophysics and Cosmology Research Unit\\
 School of Mathematical Sciences\\
 University of KwaZulu-Natal\\
 Durban 4041 \\South Africa}

\maketitle

\begin{abstract}
The condition for pressure isotropy is reduced to a recurrence
equation with variable, rational coefficients of order three. We
prove that this difference equation can be solved in general.
Consequently we can find an exact solution to the field equations
corresponding to a static spherically symmetric gravitational
potential in terms of elementary functions. The metric functions,
the energy density and the pressure are continuous and well behaved
which implies that this solution could be used to model the interior
of a relativistic sphere. The model satisfies a barotropic equation
of state in general which approximates a polytrope  close to the
stellar centre.
\end{abstract}

\section{Introduction}

Static solutions of the Einstein field equations for spherically
symmetric manifolds are important in the description of relativistic
spheres in astrophysics. The models generated may be used to
describe highly compact objects where the gravitational field is
strong as in neutron stars.  It is for this reason that many
investigators  use a variety of mathematical techniques to attain
exact solutions. On of the first models, satisfying all the physical
requirements for a neutron star, was found by Durgal and Bannerji
\cite{Durgapal}. Now there exist a number of comprehensive
collections \cite{Stephani,Skea,Lake} of static, spherically
symmetric solutions which provide a useful guide to the literature.
It is important to note that only a few of these solutions
correspond to nonsingular metric functions with a physically
acceptable energy momentum tensor.

In this paper we seek a new exact solution to the field equations
which can be used to describe the interior of a relativistic sphere.
We rewrite the Einstein equations as a new set of differential
equations which facilitates the integration process. We choose a
cubic form for one of the gravitational potentials, which we believe
has not been studied before, which enables us to simplify the
condition of pressure isotropy. This yields a third order recurrence
relation, which we manage to solve from first principles. It is then
possible to exhibit  a new exact solution to the Einstein field
equations. The curvature and thermodynamical variables appear to be
well-behaved. We also demonstrate the existence of an explicit
barotropic equation of state. For small values of the radial
coordinate close to the stellar core the equation of state
approximates a polytrope. We believe that a detailed physical
analysis of our solution is likely to lead to a realistic model for
compact objects.

\section{Static spacetimes}

Since our intention is to study relativistic stellar objects it
seems reasonable, on physical grounds, to assume that spacetime is
static and spherically symmetric. This is clearly consistent with
models utilised to study physical processes in compact objects. The
generic line element for static, spherically symmetric spacetimes is
given by
    \begin{equation}
    ds^2 = -e^{2\nu(r)}dt^2 + e^{2\lambda(r)} dr^2
    + r^{2} (d\theta^2 +\sin^{2}\theta \,d\phi^{2}) \label{spherical}
    \end{equation}
in Schwarzschild coordinates. For neutral perfect fluids the
Einstein field equations can be written in the form
\begin{subequations} \label{perfect}
    \begin{eqnarray}
\frac{1}{r^2}  [r(1-e^{-2\lambda})]'& =& \rho \label{perfect-a}
\\   -\frac{1}{r^2}\ (1-e^{-2\lambda})+\frac{2\nu'}{r}\
e^{-2\lambda}& =& p \label{perfect-b}  \\   e^{-2\lambda}
\left(\nu''+{\nu}'^{2} +\frac{\nu'}{r}  -\nu' \lambda'-
\frac{\lambda'}{r}\right) & = & p \label{perfect-c}
\end{eqnarray}
    \end{subequations}
for the line element (\ref{spherical}) where the energy density
$\rho$ and the pressure $p$ are measured relative to the comoving
fluid 4--velocity $u^a = e^{-\nu} \delta^{a}_{\,0}$ and primes
denote differentiation with respect to the radial coordinate $r.$ In
the field equations (\ref{perfect}) we are using units where the
coupling constant $\frac{8 \pi G}{c^4} = 1$ and the speed of light
is $c=1.$ An equivalent form of the field equations is obtained if
we use the transformation
\begin{equation}  A^2 y^2 (x) = e^{2 \nu(r)}, Z(x) =
e^{-2 \lambda(r)}, x = C r^2 \label{transf}
 \end{equation}
  due to
 Durgapal and Bannerji \cite{Durgapal}, where $A$ and $C$ are arbitrary constants. Under
the transformation (\ref{transf}), the system (\ref{perfect})
becomes
    \begin{subequations} \label{Finch}
    \begin{eqnarray}
    \frac{1-Z}{x}-2\dot{Z} &=&\frac{\rho}{C}            \label{Finch-a}\\
    4Z\frac{\dot{y}}{y} + \frac{Z-1}{x} &=& \frac{p}{C} \label{Finch-b}\\
    4Zx^2 \ddot{y} + 2\dot{Z}x^{2} \dot{y} + (\dot{Z}x-Z+1)y&=&0    \label{Finch-c}
    \end{eqnarray}
    \end{subequations}
where the overdot denotes differentiation with respect to the
variable $x$. Note that (\ref{Finch}) is a system of three equations
in the four unknowns $\rho, p, y$ and $Z$. The advantage of this
system lies in the fact that a solution can, upon a suitable
specification of $Z(x)$, be readily obtained by integrating
(\ref{Finch-c}) which is second order and linear in $y$.

\section{A new series solution}

A large number of exact solutions are known for the system of
equations (\ref{Finch}) that model a relativistic star with no
charge. Many of these are listed by Stephani {\it et al}
\cite{Stephani} and Finch and Skea \cite{Skea}. A comprehensive list
of static solutions, that satisfy stringent conditions for
spherically symmetric perfect fluids, was compiled by Delgaty and
Lake \cite{Lake}. The Einstein field equations in the form
(\ref{Finch}) are under-determined.
 From inspection it is clear that the simplest solutions to the system
  (\ref{Finch}) correspond to polynomials forms for $Z(x)$. As far
  as we are aware all exact solutions found previously
   correspond to forms of the gravitational potential $Z(x)$
   which are linear or quadratic in the independent variable $x$.
   Our
 approach here is to specify the gravitational potential $Z(x)$ and
attempt to solve (\ref{Finch-c}) for the potential $y.$ In an
attempt to obtain a new solution to the system (\ref{Finch}) we make
the choice
    \begin{equation}
Z=ax^3 +1   \label{ae1}
    \end{equation}
where $a$ is a constant. We suspect that the cubic form (\ref{ae1})
has not been considered before  because the resulting differential
equation in the dependent variable $y$ is difficult to solve;
quadratic forms for $Z$ are listed by Delgaty and Lake \cite{Lake}.
The quadratic form for the potential $Z$ is simpler to handle and
contains the familiar Tolman models. With the specified function
$Z,$ the condition of pressure isotropy (\ref{Finch-c}) becomes
 \begin{equation} 2(ax^3+1)\ddot{y}+3ax^2\dot{y}+axy =0.
\label{cubic4}
    \end{equation}
The linear second order differential equation (\ref{cubic4}) is
difficult to solve when $a\not=0$. We have not found a solution for
$a \neq 0$ in standard handbooks of differential equations. Software
packages such as Mathematica have also not been helpful as they
generate a solution in terms of hypergeometric functions with
complex arguments. We attempt to find a series solution to
(\ref{cubic4}) using the method of Frobenius. As the point $x=0$ is
a regular point of (\ref{cubic4}), there exist two linearly
independent solutions of the form of a power series with centre
$x=0$. We therefore can write
    \begin{equation}
y(x) = \sum^{\infty}_{n=0}c_{n} x^{n}   \label{cubic5}
    \end{equation}
where the $c_n$ are the coefficients of the series. For a legitimate
solution we need to determine the coefficients $c_n$ explicitly.

Substituting (\ref{cubic5})  into (\ref{cubic4}) yields
 \begin{eqnarray*}
4c_2+(12c_3+ac_0)x+4(6c_4+ac_1)x^2+  \\
\sum^{\infty}_{n=2}\{a[2n^2+n+1]c_n+2(n+3)(n+2)c_{n+3}\}x^{n+1}&=&0.
    \end{eqnarray*}
For this equation to hold true for all $x$ we require
    \begin{subequations}          \label{cubic9}
    \begin{eqnarray}
4c_2 &=& 0          \label{cubic9-a}        \\
12c_3 + ac_0 &=& 0  \label{cubic9-b}        \\
6c_4+ac_1   &=& 0       \label{cubic9-c}        \\
a[2n^2+n+1]c_n+2(n+3)(n+2)c_{n+3}&=&0,\, n\geq 2.   \label{cubic9-d}
    \end{eqnarray}
    \end{subequations}
Equation (\ref{cubic9-d})is a linear recurrence relation with
variable, rational coefficients of order three. General techniques
of solution for difference equations are limited to the simplest
cases and (\ref{cubic9-d}) does not fall into the known classes.
However it is possible to solve (\ref{cubic9-d}) from first
principles. Equations (\ref{cubic9-a}) and (\ref{cubic9-d}) imply
    \begin{equation}
c_2 = c_5 = c_8 = \cdots = 0.     \label{cubic11}
    \end{equation}
From (\ref{cubic9-b}) and (\ref{cubic9-d}) we generate the
expressions
    \begin{eqnarray*}
c_3 &=& -\frac{a}{2}\frac{1}{3.2} c_0   \\
c_6 &=& \frac{a^2}{2^2}\frac{2.3^2+3+1}{6.3} \frac{1}{5.2}c_0
\\   c_9 &=& -\frac{a^3}{2^3}\frac{2.6^2+6+1
}{9.6.3}\frac{2.3^2+3+1}{8.5.2}c_0.
    \end{eqnarray*}
It is clear that the coefficients $c_3, c_6, c_9, \ldots$ can all be
written in terms of the coefficient $c_0$. These coefficients
generate a pattern and we can write
\begin{equation}
c_{3n+3}=(-1)^{n+1}\left(\frac{a}{2}\right)^{n+1}\prod^{n}_{k=0}
\frac{2(3k)^2+3k+1}{(3k+3)(3k+2)}c_0
\label{cubic12}
    \end{equation}
where we have utilised the conventional symbol $\prod$ to denote
multiplication. We can obtain a similar formula for the coefficients
$c_4, c_7, c_{10}, \ldots$ From (\ref{cubic9-c}) and
(\ref{cubic9-d}) we have
    \begin{eqnarray*}
c_4&=&-\frac{a}{2}\frac{2.1^2+1+1}{4.3}c_1  \\
c_7&=& \frac{a^2}{2^2}\frac{2.4^2+4+1 }{7.4}\frac{2.1^2+1+1}{6.3}c_1
\\  c_{10}&=& -\frac{a^3}{2^3}\frac{2.7^2+7+1 }{10.7.4}
\frac{2.4^2+4+1 }{9.6.3}\frac{2.1^2+1+1}{1}c_1.
    \end{eqnarray*}
The coefficients $c_4, c_7, c_{10}, \ldots$ can all be written in
terms of the coefficient $c_1$. These coefficients generate a
pattern which is clearly of the form \begin{equation}
c_{3n+4}=(-1)^{n+1}\left(\frac{a}{2}\right)^{n+1}\prod^{n}_{k=0}
\frac{2(3k+1)^2+(3k+1)+1}{(3k+4)(3k+3)}c_1
\label{cubic14}
    \end{equation}
where $\prod$ denotes multiplication.

From (\ref{cubic11}) we observe that the coefficients $c_2, c_5,
c_8, \ldots$ all vanish. The coefficients $c_3, c_6, c_9, \ldots$
are generated from (\ref{cubic12}). The coefficients $c_4, c_7,
c_{10}, \ldots$ are generated from (\ref{cubic14}). Hence the
difference equation (\ref{cubic9-d}) has been solved and all
non-zero coefficients are expressible in terms of the leading
coefficients $c_0$ and $c_1$. We can write the series (\ref{cubic5})
as
 \begin{eqnarray}
y(x) &=&c_0+c_1x^1+c_3x^3+c_4x^4+c_6x^6+c_7x^7+c_9x^9
+c_{10}x^{10}+\cdots \nonumber \\
 &=&
c_0\left(1+\sum^{\infty}_{n=0}c_{3n+3}x^{3n+3}\right)
+c_1\left(x+\sum^{\infty}_{n=0}c_{3n+4}x^{3n+4}\right) \nonumber \\
&=& c_0\left(1+\sum^{\infty}_{n=0}(-1)^{n+1}
\left(\frac{a}{2}\right)^{n+1}\prod^{n}_{k=0}
\frac{2(3k)^2+3k+1}{(3k+3)(3k+2)}x^{3n+3}\right)+\nonumber \\   & &
c_1\left(x+\sum^{\infty}_{n=0}(-1)^{n+1}\left(\frac{a}{2}
\right)^{n+1}\prod^{n}_{k=0}\frac{2(3k+1)^2+(3k+1)+1}{(3k+4)(3k+3)}x^{3n+4}
\right) \label{cubic16}
    \end{eqnarray}
where $c_0$ and $c_1$ are arbitrary constants. Clearly
(\ref{cubic16}) is of  the form
    \begin{equation}
y(x)=c_0y_1(x)+c_1y_2(x)        \label{cubic17}
    \end{equation}
where
    \begin{eqnarray*}
y_1(x) &=&\left(1+\sum^{\infty}_{n=0}(-1)^{n+1}
\left(\frac{a}{2}\right)^{n+1}\prod^{n}_{k=0}
\frac{2(3k)^2+3k+1}{(3k+3)(3k+2)}x^{3n+3}\right) \\ \\
y_2(x)
&=&\left(x+\sum^{\infty}_{n=0}(-1)^{n+1}
\left(\frac{a}{2}\right)^{n+1}\prod^{n}_{k=0}
\frac{2(3k+1)^2+(3k+1)+1}{(3k+4)(3k+3)}x^{3n+4}
\right) \end{eqnarray*}
are linearly independent solutions of (\ref{cubic4}). Therefore we
have found the general solution to the differential equation
(\ref{cubic4}) for the particular gravitational potential $Z$ given
in (\ref{ae1}). The advantage of the solutions in (\ref{cubic17}) is
that they are expressed in terms of a series with real arguments
unlike the complex arguments given by software packages.

\section{Physical models}

From (\ref{cubic17}) and the Einstein field equations (\ref{Finch})
we generate the exact solution
    \begin{subequations}      \label{cubic19}
    \begin{eqnarray}
e^{2\lambda}&=&\frac{1}{ax^3+1} \label{cubic19-a}   \\
e^{2\nu}&=& A^2 y^2         \label{cubic19-b}   \\
\frac{\rho}{C}&=& -7ax^2        \label{cubic19-c}   \\
\frac{p}{C}&=& 4(ax^3+1)\frac{\dot{y}}{y}+ax^2  \label{cubic19-d}
    \end{eqnarray}
    \end{subequations}
In the above the  quantity $y$ is given by (\ref{cubic17}), $x=Cr^2$
and $a$ is a constant. This solution has a simple form and is
expressed completely in terms of elementary functions. The
expressions given above have the advantage of simplifying the
analysis of the physical features of the solution, and will assist
in the description of a relativistic compact bodies such as neutron
stars.

Consider a relativistic sphere where $0 \leq x \leq b$ where
$b=CR^2$ and $R$ is the stellar radius. We note that the functions
$\nu$ and $\lambda$ have constant values at the centre $x=0$. The
function $\rho$ vanishes at the centre. The pressure $p$ has a
constant value at $x=0$. Hence the gravitational potentials and the
matter variables are finite at the centre. Since
$y(x)=c_0y_1(x)+c_1y_2(x)$ is a well defined series on the interval
$[0, b]$ the quantities $\nu$, $\lambda$, $\rho$ and $p$ are
nonsingular and continuous. If $a<0$ then the energy density $\rho
>0$. The constants $c_0$ and $c_1$ can be chosen such that the
pressure $p>0$. Consequently the energy density and the pressure are
positive on the interval $[0,b]$. At the boundary $x=b$ we must have
\[ e^{-2\lambda (R)} = aC^3R^6 +1 = 1- \frac{2M}{R} \]
for a sphere of mass $M$; this ensures  that the interior spacetime
matches smoothly to the Schwarzschild exterior. For the speed of
sound to be less than the speed of light we require that
\[ 0 \leq \frac{dp}{d\rho} \leq 1 \]
in our units. This inequality will constrain the values of the
constants $a$, $c_0$, $c_1$, $A$ and $C$. From this qualitative
analysis we believe that the solution found can be used as a basis
to describe realistic relativistic stars. We believe that a detailed
physical analysis is likely to lead to realistic models for compact
objects.

Our solution has the interesting feature of admitting an explicit
barotropic equation of state. We observe from (\ref{cubic19-c}) that
    \[x=\sqrt{\frac{\rho}{-7aC}}, \quad a<0 \]
and the variable $x$ can be written in terms of $\rho$ only. The
function $y$ in (\ref{cubic17}) can be expressed in terms of $\rho$
and the variable $x$ is eliminated. Consequently the pressure $p$ in
(\ref{cubic19-d}) is expressible in terms of $\rho$ only, and we can
write
    \[p=p(\rho). \]
Thus the solution in (\ref{cubic19}) obeys a barotropic equation of
state. This highly desirable feature is unusual for most exact
solutions as pointed out in Stephani {\it et al} \cite{Stephani}.
For small values of $x$ close to the stellar centre we have
 $y \approx c_0 + c_1x$. Then from (\ref{cubic19-c})
 and (\ref{cubic19-d}) we have the approximation
 \begin{equation}
 \frac{p}{C} \approx \frac{4 c_1}{c_0 + c_1 \sqrt{\frac{\rho}{-7aC}}}
 \label{polytropic}
 \end{equation}
Therefore  for small values of $x$ close to stellar centre
(\ref{polytropic}) implies that we have the approximate equation of
state
\[ p \propto \rho^{-1/2} \]
which is of  the form of a polytrope.

We point out that the solutions presented in this paper may be
extended to anisotropic matter.  In recent years a number of
researchers  have proposed models corresponding to anisotropic
matter where the radial component of the pressure differs from the
angular component. The physical motivation for the analysis of
anisotropic matter is that anisotropy  affects the critical mass,
critical surface redshift and stability of highly compact bodies.
These investigations are contained in the papers
\cite{ChaisiMaharajA,DevGleiser,DevGleiser2,HerreraMartin,
HerreraTroconis,Ivanov,MakHarko2002,MakHarko}, among others. It
appears that anisotropy may be important in fully understanding the
gravitational behaviour of boson stars and the role of strange
matter with densities higher than neutron stars. Mak and Harko
\cite{MakHarko2002} and Sharma and Mukherjee
\cite{SharmaMukherjee2002} have observed that anisotropy is a
crucial ingredient in the analysis of dense stars with strange
matter. The simple form of our solutions allows for extension to
study such matter by adapting the energy momentum tensor to include
both radial and tangential pressures.

\section*{Acknowledgements}
AJ is grateful to the University of KwaZulu-Natal for a scholarship.
AJ and SDM thank the National Research Foundation of South Africa
for financial support.

\end{document}